\documentclass[conference,letterpaper]{IEEEtran}

\usepackage{pdfpages,multirow}
\usepackage{amssymb,enumitem}
\usepackage[hidelinks]{hyperref}
\hypersetup{
    colorlinks=true,
    linkcolor=blue,
    filecolor=magenta,      
    urlcolor=cyan,
    pdftitle={Overleaf Example},
    pdfpagemode=FullScreen,
}
\graphicspath{ {./images/} }

\usepackage[utf8]{inputenc} 
\usepackage[T1]{fontenc}
\usepackage{url}
\usepackage{ifthen}
\usepackage{cite}
\usepackage[cmex10]{amsmath} 
\usepackage{mleftright}
\mleftright            
\usepackage{tikz}
\usepackage{amsfonts}
\usepackage{graphicx}         % provides \includegraphics{...} to
\usepackage{booktabs}         % fixes poor spacing in tables and
\usepackage{amsthm}

\newtheorem*{rep@theorem}{\rep@title}
\newcommand{\newreptheorem}[2]{%
\newenvironment{rep#1}[1]{%
 \def\rep@title{#2 \ref{##1}}%
 \begin{rep@theorem}}%
 {\end{rep@theorem}}}
\makeatother

\newreptheorem{theorem}{Theorem}
\newreptheorem{lemma}{Lemma}
\newreptheorem{proposition}{Proposition}

\newtheorem{theorem}{Theorem}
\newtheorem{proposition}{Proposition}

\newtheorem{remark}{Remark}

\theoremstyle{definition}
\definecolor{britishracinggreen}{rgb}{0.0, 0.26, 0.15}
\newtheorem{definition}{Definition}

\title{{Improved Probabilistic Lower Bounds for Separable Matrices}}

\author{
  \IEEEauthorblockN{Daniil Goshkoder}
  \IEEEauthorblockA{Skolkovo Institute of Science and Technology \\
                    Moscow, Russia \\
                    Email: daniilgoshkoder@mail.ru}
  \and
  \IEEEauthorblockN{Nikita Polyanskii and Ilya Vorobyev}
  \IEEEauthorblockA{IOTA Foundation\\ 
                    Berlin, Germany\\
                    Email: \{nikita.polyansky@gmail.com, vorobyev.i.v@yandex.ru\}}
}

\begin{document}
\maketitle

\begin{abstract}
    This work focuses on non-adaptive combinatorial group testing, with a primary goal of efficiently identifying a set of at most $d$ defective elements among a given set of $n$ elements using the fewest possible tests. Non-adaptive combinatorial group testing often employs disjunctive matrices (DM) and separable matrices (SM). This paper discusses separable matrices and recently introduced list-decoding separable matrices (LDSM) with list size $n^{1/d}$, which allow for non-adaptive identification of defectives with the decoding complexity linear in the number of tests and the number of elements. In our study, we distinguish two subclasses of these matrices: matrices which can be used when the number of defectives $d$ is a priori known ($d$-SM and $(d, n^{1/d})$-LDSM), and matrices which can be used for any subset of at most $d$ defectives ($\bar{d}$-SM and $(\bar{d}, n^{1/d})$-LDSM). Our contribution lies in deriving new lower bounds on the rates of $d$-SM, $\bar{d}$-SM, $(d, n^{1/d})$-LDSM and $(\bar{d}, n^{1/d})$-LDSM for an arbitrary number $d \ge 3$ of defectives.
\end{abstract}

\section{Introduction} \label{sec: introduction}

The concept of group testing is a widely recognized combinatorial problem that Dorfman first introduced in 1943~\cite{dorfman1943detection}. In this scenario, we are given a large set of $n$ elements, with some of them being \textit{defective}. The objective is to identify the set of defective elements using the least possible number of \textit{tests}, where each test involves examining a carefully selected subset of potential defectives. The outcome of a test is considered \textit{positive} if at least one element of the subset is defective and \textit{negative} if none of the elements are defective.

In recent years, group testing has gained significant attention and has been employed in various domains. Its applications include combating the Covid-19 pandemic~\cite{aldridge2022pooled}, chemical and electrical testing, coding, drug screening, multiaccess channel management, data verification, data science, and more~\cite{du2000combinatorial, aldridge2019group}.

Two main types of algorithms are used in group testing. The first type is \textit{adaptive} algorithms, where tests are conducted one at a time, and the results of each test are used to inform the design of subsequent tests. The second type is \textit{non-adaptive} algorithms, where all tests are predefined and can be executed simultaneously or in parallel.

Group testing problems can also be categorized into two models: \textit{combinatorial} and \textit{probabilistic}. In the combinatorial model, the algorithms always identify the set of defectives correctly, given that its size is at most $d$. In the probabilistic model, there is a probabilistic distribution on the sets of defectives, and a decoding algorithm identifies the set of defectives with a high probability.

In this paper, we primarily focus on non-adaptive combinatorial algorithms. For a broader perspective on the field, we refer to surveys such as~\cite{du2000combinatorial, aldridge2019group, d2014lectures, cicalese2013fault}, which provide comprehensive overviews of different models and approaches. In addition, we mention works addressing the problems of estimating the number of defectives~\cite{cheng2011efficient, falahatgar2016estimating, d2019hypothesis}, the identification of a sufficiently large subset of non-defectives~\cite{sharma2017computationally, sharma2018finding}, and more recent approaches that enhance group testing methods by incorporating side information \cite{cao2023group}.

\subsection{Related works}

Any non-adaptive combinatorial group testing (CGT) scheme with $n$ samples and $t$ tests can be described using a matrix $C$ of size $t \times n$. Each row of $C$ represents a specific test, and each column represents a sample. In this matrix, the intersection of the $i$-th column and the $j$-th row contains a value $1$ iff the sample $i$ is used in the test $j$. The outcome (result vector) of the test scheme defined by $C$ is a Boolean sum of the columns corresponding to the elements from the defective set.

In a landmark paper by Kautz and Singleton in 1964~\cite{kautz1964nonrandom}, two families of matrices were introduced to address non-adaptive CGT problems. The first family, called \textit{$d$-separable matrices} (or briefly $d$-SM)\footnote{{Such matrices are called \textit{uniquely decipherable} codes in~\cite{kautz1964nonrandom} and \textit{union-free} codes and families in~\cite{erdos1982families, erdHos1985families, coppersmith1998new, vorobyev2021fast}.}}, ensures that different sets of {not more than $d$} defectives yield unique test result vectors. This property allows us to reconstruct the set of defectives from the test results. However, apart from brute force, which involves exhaustive search, no other decoding algorithm for $d$-SM is known. This decoding scheme has a worst-case time complexity of $O(tn^d)$, where $t$ is the number of tests and $d$ is the maximum possible number of defectives. The second family of matrices, known as the \textit{disjunctive matrices} (or briefly $d$-DM), offers {a more efficient decoding algorithm with a time complexity of $O(tn)$, which finds all columns of the matrix covered by the result vector.} {It is also known that $d$-disjunctive matrices form a subfamily of $d$-separable matrices. This means that the minimum number of tests to detect defectives for $d$-SM does not exceed the minimum number of tests for $d$-DM.} {The generalization of disjunctive matrices for list-decoding, called \textit{$d$-disjunctive list-decoding matrices}, is introduced and discussed in~\cite{dyachkov1983survey, d2014lectures}. Note that $d$-disjunctive list-decoding matrices cannot be independently used to solve the problem of non-adaptive CGT, but they can be used in a two-stage testing scheme, see, for example~\cite{vilenkin1998constructions}. The authors of~\cite{de2005optimal} present two-stage algorithms for combinatorial and probabilistic GT problems using another generalization of disjunctive codes called selectors.}

{The study of lower and upper bounds on the largest rates\footnote{Here and below, largest rate refers to the maximum asymptotic rate $\limsup \frac{\log_2 n}{t}$ with $t \to \infty$, unless otherwise stated.} of the described constructions began with the case of $d=2$. The authors of~\cite{coppersmith1998new} presented a lower bound of $0.3135$ on the rate of $2$-SM, which is significantly better than the best-known lower bound of $0.1825$~\cite{dyachkov1989superimposed} and only slightly less than the best-known upper bound of $0.3219$~\cite{d1982bounds} on the rate of $2$-DM. In~\cite{dyachkov1989superimposed, d2014lectures}, the authors presented lower and upper bounds on the rates of SM and DM for the case of an arbitrary number of defectives $d \ge 2$, as well as for the case of $d \to \infty$.
% In~\cite{dyachkov1989superimposed, d2014lectures}, the results of the case $d=2$ were partially generalized to the case of an arbitrary number of defectives $d \ge 2$, as well as for the case of $d \to \infty$.
Note that lower bounds for SM for the cases of fixed $d$ and $d\to\infty$ are published without proof in~\cite{d2014lectures}, and these bounds work for a special class of SM which assumes that the number $d$ of defectives is a priori known. For the problem of exactly $d$ defectives with $d \to \infty$, the upper bounds of $\frac{2 \log_2 d}{d^2}(1 + o(1))$ for DM and $\frac{4 \log_2 d}{d^2}(1 + o(1))$ for SM are also shown in~\cite{d2014lectures}. At the same time, there are no special lower bounds on the rate of SM which differ from the best-known lower bounds on the rates of other constructions for the case when the number of defectives does not exceed $d$ for $d \ge 3$.} {A more detailed relationship between DM and SM, and their asymptotic largest rates in the case of $d\to\infty$ was considered in~\cite{chen2007exploring}. In particular, the authors show that for any $d$ we can consider a $2d$-SM with $n$ columns and $t$ rows for the CGT problem of exactly $2d$ defectives, add at most one row to this matrix and obtain a $d$-DM.}

% {As part of further research in this area, the authors investigate the problem of constructing new designs to solve the problem of non-adaptive CGT, combining the advantages of DM and SM: constructions with $O(tn)$ decoding complexity and a greater rate than for DM.}

{The first effort to combine the advantages of two known constructions of DM and SM was proposed in~\cite{fan2021strongly}. The authors introduce a new notion of \textit{strongly $d$-separable matrix} (or briefly $d$-SSM) for non-adaptive CGT. It was proved that the classes of SSM for the CGT problems with exactly $d$ defectives and not more than $d$ defectives coincide. The relationship between DM, SM and SSM was also described in~\cite{fan2021strongly}: a $d$-DM is a $d$-SSM, a $d$-SSM is a $d$-SM, and a $d$-SSM decoding complexity is $O(tn)$ as for $d$-DM.} Additionally, in~\cite{qian2022improved}, the authors show a lower bound of $0.2237$ on the rate of $2$-SSM and a lower bound of $0.0974$ on the rate of $3$-SSM against the lower bounds of $0.1825$ for $2$-DM and $0.079$ for $3$-DM~\cite{dyachkov1989superimposed}. {These results show that for $d=2,3$ a family of $d$-SSM may require less number of tests than a family of $d$-DM for solving the problem of non-adaptive CGT.}

{In~\cite{vorobyev2021fast}, the authors modify SM to provide them with an efficient $O(tn)$ decoding algorithm. The new family called \textit{$d$-union-free matrices with fast decoding} (or briefly $d$-UFFDM) is a subfamily of separable matrices with an additional constraint. For the introduced subclass, at the first decoding step, an element is considered \textit{potentially defective} if, for all tests that include this element, the test result is positive. The additional constraint states that the number of potentially defective elements is at most $n^{1/d}$. Next, we need to do an exhaustive search over all possible potentially defective sets of not more than $d$ elements. Since the number of potentially defectives is upper bounded by $n^{1/d}$, the decoding complexity does not exceed $O(tn + t\cdot (n^{\frac{1}{d}})^d) = O(tn)$. The authors of~\cite{vorobyev2021fast} also show that a $d$-DM is a $d$-UFFDM, and a $d$-UFFDM is a $d$-SM. Moreover, the authors present a lower bound of $0.3017$ on the largest rate of the new family of matrices for the case of $d=2$ defectives. This lower bound is considerably better than the lower bounds achieved by $2$-DM and $2$-SSM, and only slightly less than the lower bound of $0.3135$ proved for $d$-SM in~\cite{coppersmith1998new}.}

{Later\footnote{{The authors point out that the research in~\cite{fan2024list} was conducted independently and concurrently with the research presented in this paper.}}, the authors of~\cite{fan2024list} introduce the generalization of UFFDM contruction, called this generalization \textit{$(d, L)$-list-decoding separable matrices} (or briefly $(d, L)$-LDSM), which coincide with $d$-UFFDM family when $L=n^{1/d}$. Moreover, it was shown that $(d, d)$-LDSM family coincide with $d$-DM family and $(d, n)$-LDSM family coincide with $d$-SM family. In general, the decoding complexity of $(d, L)$-LDSM is $O(\max\{tn, tL^d\})$. Note that for $L=n^{1/d}$, the decoding complexity is $O(tn)$ as for $d$-DM. In~\cite{fan2024list}, the authors also present the lower bounds on the largest rate of the introduced construction for any $d \ge 3$ using the probabilistic method. For small values of $d$, the presented lower bounds on the rate of $(d, n^{1/d})$-LDSM are better than the best-known lower bounds on the largest rates of $d$-DM and $d$-SSM, which have the same decoding complexity of $O(tn)$.}

In papers~\cite{macula2004trivial, mazumdar2012almost, d2015almost, malyutov1978separating, sebHo1985two, zhigljavsky2003probabilistic, aldridge2017almost}, a probabilistic model of group testing is considered. The authors of these papers discuss $\epsilon$-almost disjunctive matrices and $\epsilon$-almost separable matrices. These families of matrices are used to identify the set of exactly $d$ defectives with error probability not greater than $\epsilon$ for any $\epsilon > 0$. {As in the combinatorial case, the best known decoding complexity is $O(tn)$ for $\epsilon$-almost $d$-DM and $O(tn^d)$ for $\epsilon$-almost $d$-SM.} {Using the probabilistic method, the authors of~\cite{d2015almost} present the best-known lower bound on the rate of $\epsilon$-almost $d$-DM for finite $d$ and show that this rate is at least $\frac{\ln 2}{d}(1+o(1))$ for $d\to\infty$, i.e., the minimum number of tests is $t = \frac{d \log_2 n}{\ln 2} (1 + o(1))$ for large $d\ll n$. In~\cite{freidlina1975design}, the authors show that $\epsilon$-almost $d$-SM exists with $t=(d+o(1))\cdot\log_2 n$ for fixed $d$, i.e., this is a rate optimal construction. The authors of~\cite{aldridge2017almost, mazumdar2012almost} analyze $\epsilon$-almost $d$-disjunctive matrices and $\epsilon$-almost separable matrices with $d=o(n)$.}

In the paper~\cite{inan2019optimality}, the authors show that the Kautz and Singleton construction~\cite{kautz1964nonrandom} achieves an order-optimal number of tests $O(d\log n)$ for $d=\Omega(\log^2 n)$. Moreover, from the results of paper~\cite{bassalygo2013multiple} it follows that the same construction is order-optimal for $d=(c+o(1))\log_2 n$ for any $c\in (0, \ln 2)$. In~\cite{coja2020optimal}, the authors present a group testing scheme with an asymptotically optimal number of tests and a polynomial in $n$ decoding algorithm. In~\cite{cai2017efficient, inan2019optimality, lee2019saffron, price2020fast, cheraghchi2020combinatorial, bondorf2020sublinear, wang2023quickly}, algorithms are being built that have a sublinear decoding complexity in terms of the number of elements $n$, but these constructions require more tests. However, we do not compare the results of our work with the results from~\cite{cai2017efficient, inan2019optimality, lee2019saffron, price2020fast, cheraghchi2020combinatorial, coja2020optimal, bondorf2020sublinear, wang2023quickly, bassalygo2013multiple}, as their results are obtained for the non-constant number of defectives $d$.

\subsection{Our contribution}

In our paper, we work with $d$-SM and $(d, n^{1/d})$-LDSM to identify a set of exactly $d$ defectives, and $\bar{d}$-SM and $(\bar{d}, n^{1/d})$-LDSM to identify a set of not more than $d$ defectives among $n$ samples with $t$ tests. Here and below, we use the notations of $d$ and $\bar{d}$ to separate these cases. With the help of probabilistic methods, we improve and supplement the results of~\cite{d2014lectures} for SM and the results of~\cite{vorobyev2021fast, fan2024list} for LDSM with list size $L=n^{1/d}$. Our contribution is summarized below:
\begin{itemize}[wide, labelwidth=1pt, labelindent=3pt]
    \item For an arbitrary number $d\ge 3$ of defectives, we present analytical lower bounds on the rates of $d$-SM and $(d, n^{1/d})$-LDSM. We show that the presented lower bounds coincide for small values of $d$.
    \item We show how to use the lower bounds on the rate of $d$-SM and $(d, n^{1/d})$-LDSM to derive lower bounds on the rate of $\bar{d}$-SM and $(\bar{d}, n^{1/d})$-LDSM. For small values of $d$, the derived lower bounds for $\bar{d}$-SM and $(\bar{d}, n^{1/d})$-LDSM coincide and outperform the known lower bounds on the rates of $d$-DM~\cite{dyachkov1989superimposed} and $d$-SSM~\cite{qian2022improved}, e.g., see numerical values in Table \ref{tab: comparison}. Recall that $(\bar{d}, n^{1/d})$-LDSM, $d$-DM and $d$-SSM have the same decoding complexity of $O(tn)$ and allow identifying any subset of at most $d$ defectives.
\end{itemize}

A comparison of all obtained lower bounds with the previously known lower bounds for $d$-separable matrices, $\bar{d}$-separable matrices, $d$-disjunctive matrices, $d$-strongly separable matrices and $(\bar{d}, L)$-list-decoding separable matrices is shown in Table \ref{tab: comparison}. We highlight in bold the results obtained in this work.

\begin{table*}[htbp]
  \centering
    \caption{Lower bounds on the rate and decoding complexity of group testing matrices.}
    \label{tab: comparison}
    \begin{tabular}{| c | c | c | c | c | c || c |}
        \hline
        $d$ & 2 & 3 & 4 & 5 & 6 & Decoding Complexity \\ 
        \hline
        \hline
        $d$-DM & 0.183~\cite{dyachkov1989superimposed} & 0.079~\cite{dyachkov1989superimposed} & 0.044~\cite{dyachkov1989superimposed} & 0.028~\cite{dyachkov1989superimposed} & 0.019~\cite{dyachkov1989superimposed} & $O(tn)$ \\
        \hline
        $d$-SSM & 0.224~\cite{qian2022improved} & 0.097~\cite{qian2022improved} & 0.044~\cite{dyachkov1989superimposed, fan2021strongly} & 0.028~\cite{dyachkov1989superimposed, fan2021strongly} & 0.019~\cite{dyachkov1989superimposed, fan2021strongly} & $O(tn)$ \\
        \hline
        $(\bar{d}, n^{1/d})$-LDSM & 0.302~\cite{vorobyev2021fast} & \textbf{0.142} & \textbf{0.079} & \textbf{0.044} & \textbf{0.028} & $O(tn)$ \\
        \hline
        $\bar{d}$-SM & 0.314~\cite{coppersmith1998new} & \textbf{0.142} & \textbf{0.079} & \textbf{0.044} & \textbf{0.028} & $O(tn^d)$ \\
        \hline
        $(d, n^{1/d})$-LDSM & 0.302~\cite{vorobyev2021fast} & \textbf{0.142} & \textbf{0.082} & \textbf{0.053} & \textbf{0.037} & $O(tn)$ \\
        \hline
        $d$-SM & 0.314~\cite{coppersmith1998new} & \textbf{0.142} & \textbf{0.082} & \textbf{0.053} & \textbf{0.037} & $O(tn^d)$ \\
        \hline
    \end{tabular}
\end{table*}

\subsection{Applications}
The study of disjunctive and separable matrices is motivated not only by their intrinsic combinatorial interest but also by a wide range of applications in signal processing and communication systems. In particular, the group testing approaches naturally arizes in models where multiple input signals are superimposed and one aims to efficiently identify a small subset of active sources. Classical examples include multiuser detection in \textit{multiple access channels}, where users transmit over shared resources and the receiver must detect the active set with minimal overhead. Disjunctive and separable matrix families provide the mathematical foundation for designing robust non-adaptive access schemes, enabling unique identification of active users.

Another important application of our research can be described in terms of \textit{sparse signal models}. A signal model is called sparse if the number of active components contributing to the observed outcome (the set of defectives) is relatively small. The study of such models has recently attracted significant attention, driven by applications in social and computer networks, transportation systems, and power-line monitoring. Since most real-world systems are inherently large-scale but sparse, special models have been developed to capture and analyze this structure. Disjunctive and separable matrices naturally serve as measurement designs for such tasks, enabling efficient identification of sparse activity patterns in high-dimensional signal processing problems.

An additional aspect that deserves special attention is the decoding complexity of the matrix families under consideration. For a matrices of size $t \times n$, the best known decoding complexity for classical $\bar{d}$-separable matrices is of order $O(tn^{d})$, whereas $d$-disjunctive and $(\bar{d}, n^{1/d})$-list-decoding separable matrices admit more efficient algorithms with complexity $O(tn)$. By reducing computational cost, list-decoding separable matrices become well suited for large-scale applications, while the obtained lower bounds on their rates establish a theoretical foundation with potential applications for high-dimensional signal processing.

\subsection{Paper organization}
{The rest of the paper is organized as follows. In Section \ref{sec: preliminaries}, we introduce important definitions and notations that are used throughout the paper. We formulate Theorems \ref{th: d-SM lower bound} and \ref{th: d-LDSM lower bound} with the lower bounds on the rates of $d$-SM and $(d, n^{1/d})$-LDSM in Section \ref{sec: SM LDSM lower bound}.In in Section \ref{sec: <= d SM LDSM} we describe the lower bounds on the rates of $\bar{d}$-SM and $(\bar{d}, n^{1/d})$-LDSM. Section \ref{sec: conclusion} concludes the main part of the paper.} We delegate proofs of the main theorems and some additional information to Appendices \ref{sec: appendix a}, and \ref{sec: appendix b}.

\section{Preliminaries} \label{sec: preliminaries}
Let $[n]$ denote the set of integers $\{1,2, \ldots, n\}$. The \textit{weight} of a vector is defined as the number of non-zero entries it contains. {Define an operation of \textit{Boolean sum} of binary numbers as $x \bigvee y = 0$ when $x=y=0$ and $1$ otherwise.} A vector $\boldsymbol{z}=(\boldsymbol{z}(1), \ldots, \boldsymbol{z}(n))$ is called a \textit{Boolean sum} of vectors $\boldsymbol{x}=(\boldsymbol{x}(1), \ldots, \boldsymbol{x}(n))$ and $\boldsymbol{y}=(\boldsymbol{y}(1), \ldots, \boldsymbol{y}(n))$ if for each $i$, $\boldsymbol{z}(i)=\boldsymbol{x}(i) \vee \boldsymbol{y}(i)$. This operation is denoted as $\boldsymbol{z}=\boldsymbol{x} \bigvee \boldsymbol{y}$. Say that a vector $\boldsymbol{x}$ \textit{covers} a vector $\boldsymbol{y}$ if $\boldsymbol{x} \bigvee \boldsymbol{y}=\boldsymbol{x}$. The \textit{binary entropy} function $h(x)$ is defined as $h(x)=-x \log_2 x-(1-x) \log_2 (1-x)$.

Let us consider a non-adaptive group testing algorithm that consists of $t$ tests. This algorithm can be described using a binary matrix $C$ with size $t \times n$, in which each row represents a specific test and each column represents a sample. In this matrix, the intersection of column $i$ and row $j$ contains a value $1$ if the sample $i$ is included in the test $j$, and $0$ otherwise. We can refer to the columns of the matrix $C$ as $\boldsymbol{c}_1, \ldots, \boldsymbol{c}_n$. To express the testing outcomes, we use a vector $\boldsymbol{r}$ defined as follows
\begin{equation*}
    \boldsymbol{r}=\bigvee_{i \in D} \boldsymbol{c}_i,
\end{equation*}
where $D$ denotes the set of defectives. If the set $D$ is empty, we define the vector $\boldsymbol{r}$ as an all-zero vector.

\subsection{Combinatorial GT problem}

{We start with describing known constructions applicable to solving the CGT problem.} Let us introduce the definitions of the separable matrices and the disjunctive matrices, which were first presented in the original work~\cite{kautz1964nonrandom} as uniquely decipherable and zero-drop rate codes. The well-known connection between these two families of matrices was also established in~\cite{kautz1964nonrandom}.

\begin{definition}\label{def: d-separable matrix}
    Call a $t \times n$ matrix $C$ a \textit{$d$-separable} (or \textit{$\bar{d}$-separable}) matrix if for any two different sets $D_1, D_2 \subset[n]$, $\left|D_i\right| = d$ (or $\left|D_i\right| \le d$) for $i=1,2$, the Boolean sums $\bigvee_{i \in D_1} \boldsymbol{c}_i$ and $\bigvee_{i \in D_2} \boldsymbol{c}_i$ are distinct.
\end{definition}

\begin{definition}\label{def: d-disjunctive matrix}
    Call a $t \times n$ matrix $C$ a \textit{$d$-disjunctive matrix} if for any set $D \subset[n], |D|=d$, the Boolean sum $\bigvee_{i \in D} c_i$ doesn't cover any column $\boldsymbol{c}_j$ for $j \in[n] \backslash D$.
\end{definition}

\begin{proposition}\label{prop: SM DM connectness}
    A $d$-DM is a $\bar{d}$-SM. A $\bar{d}$-SM is a $(d-1)$-DM.
\end{proposition}

\begin{remark}
    {Separable matrices are also called \textit{union-free codes} or \textit{union-free families} in literature, see~\cite{coppersmith1998new, vorobyev2021fast}. In this paper, we prefer the notation of separable matrices, and for simplicity, we use the abbreviations of $d$-SM and $d$-DM for separable matrices and disjunctive matrices, respectively.}
\end{remark}

Separable matrices (SM) have applications in non-adaptive CGT for identifying a set of at most $d$ or exactly $d$ defectives. The concept of SM guarantees the existence of a unique set $D$, with cardinality at most $d$ or exactly $d$, that can reproduce the observed outcome vector $\boldsymbol{r}$. However, it is unknown if there is a more efficient decoding algorithm than iterating through all possible sets of size at most $d$ or exactly $d$, resulting in a computational complexity of $O\left(t n^d\right)$.

Disjunctive matrices (DM) can also be used in non-adaptive CGT for finding a set of at most $d$ defectives. However, the decoding process can be performed more efficiently with a complexity of $O(tn)$. The decoding algorithm produces a list of indices $i$ for which the column $\boldsymbol{c}_i$ is covered by the outcome vector $\boldsymbol{r}$. In simpler terms, the algorithm identifies all the elements that only appear in positive tests. Consequently, all defective elements are guaranteed to be detected without including any other elements. This decoding algorithm is commonly known as Combinatorial Orthogonal Matching Pursuit (COMP) in the context of probabilistic group testing~\cite{aldridge2019group}.

{Let us recall the definition of another design that allows us to solve the non-adaptive CGT problem: strongly separable matrices, introduced in~\cite{fan2021strongly}.}

\begin{definition}\label{def: d-strongly separable matrix}
    A binary $t \times n$ matrix $C$ is called a \textit{strongly $d$-separable matrix} ($d$-SSM), if for any $D_0 \subset[n],\left|D_0\right|=d$, we have
    \begin{equation*}
    \bigcap_{D^{\prime} \in U\left(D_0\right)} D^{\prime}=D_0,
    \end{equation*}
    where
    \begin{equation*}
    U\left(D_0\right)=\left\{D^{\prime} \subset[n]: \bigvee_{i \in D^{\prime}} \boldsymbol{c}_i=\bigvee_{i \in D_0} \boldsymbol{c}_i\right\}.
    \end{equation*}
\end{definition}

\begin{remark}
    {We call a $t \times n$ matrix $C$ a $\bar{d}$-SSM if the condition $|D_0|=d$ in Definition \ref{def: d-strongly separable matrix} is replaced by $1 \le \left|D_0\right| \le d$.}
\end{remark}

Note that the set $U(D_0)$ may include subsets of columns of any cardinality, not only $d$. {In~\cite{fan2021strongly}, the authors show that definitions of $d$-SSM and $\bar{d}$-SSM coincide and that $d$-SSM lie in between $\bar{d}$-separable matrices and $d$-disjunctive matrices.}

\begin{proposition}\label{prop: <=d-ssm =d-ssm}{\cite{fan2021strongly}}
A matrix $C$ is $d$-SSM iff $C$ is $\bar{d}$-SSM.
\end{proposition}

\begin{proposition}\label{prop: ssm sm dm connect}{\cite{fan2021strongly}}
A $d$-DM is a $d$-SSM. A $d$-SSM is a $\bar{d}$-SM.
\end{proposition}

For $d$-SSM, there is an efficient $O(tn)$ decoding algorithm, presented in the paper~\cite{fan2021strongly}. Such an algorithm coincides with the DD (definitely defective) algorithm, which has been previously used in probabilistic group testing~\cite{aldridge2019group, aldridge2014group}.

{Move on to the description of the $(d, L)$-list-decoding separable matrices. The concept of utilizing separable matrices within a list-decoding algorithm was first proposed in~\cite{vorobyev2021fast}. The authors of~\cite{vorobyev2021fast} introduced the so-called union-free (separable) matrices with fast decoding, for which the list-decoding with a list size of $L = n^{1/d}$ achieves a decoding complexity of $O(tn)$. This construction was later generalized in~\cite{fan2024list}, where the authors introduced the $(d, L)$-list-decoding separable matrices.}

{\begin{definition}\label{def: d-LDSM}
A $t \times n$ binary matrix $C$ is a \textit{$(d, L)$-list-decoding separable} (or \textit{$(\bar{d}, L)$-list-decoding separable}) matrix, if the following two conditions hold:
\begin{enumerate}
\item Matrix $C$ is a $d$-separable (or $\bar{d}$-separable) matrix.
\item Each possible vector of outcomes $\boldsymbol{r}$, i.e., a vector, which can be obtained as a Boolean sum of exactly (at most) $d$ columns of $C$, covers at most $L$ columns from $C$.
\end{enumerate}
\end{definition}}

{For simplicity, we denote list-decoding separable matrices as $(d, L)$-LDSM and $(\bar{d}, L)$-LDSM. In~\cite{fan2024list} it was shown that the family of $(\bar{d}, d)$-LDSM coincide with the family of $d$-DM, and the family of $(\bar{d}, n)$-LDSM coincide with the family of $\bar{d}$-SM. In other words, the construction of $(\bar{d}, L)$ is also a generalisation of two classic CGT constructions.}

{In~\cite{fan2024list}, the authors present the decoding algorithm for $(\bar{d}, L)$-LDSM and $(d, L)$-LDSM. This algorithm consists of two steps.}

\textit{Step 1}. Find {potentially defective} set $\hat{D}$ of all indexes $i$, such that, $\boldsymbol{c}_i$ is covered by the vector of outcomes $\boldsymbol{r}$.

\textit{Step 2}. Iterate over all sets $D_0 \subset \hat{D}$, $\left|D_0\right| \le d$ (or $\left|D_0\right| = d$). If $\bigvee_{i \in D_0} \boldsymbol{c}_i = \boldsymbol{r}$, then output $D_0$ as an answer and finish the algorithm.

{From Definition \ref{def: d-LDSM} it follows that the number of potentially defective elements after the first decoding step is upper bounded by $L$. Therefore, the decoding complexity does not exceed $O(tn + tL^d) = O(\max\{tn, tL^d\})$.}

\begin{proposition}\label{prop: ldsm algorithm correctness}{\cite{fan2024list}}
The described algorithm successfully finds the set of at most (exactly) $d$ defectives. The number of required operations is $O(\max\{tn, tL^d\})$.
\end{proposition}

{\begin{remark}
With $L=n^{1/d}$ the decoding complexity of $(\bar{d}, n^{1/d})$-LDSM is $O(tn)$, as for $d$-DM and $d$-SSM. This result and the same decoding algorithm for $L=n^{1/d}$ were previously presented in~\cite{vorobyev2021fast}. In the following part of the paper, we focus mainly on the case of $L=n^{1/d}$, i.e., on the constructions of $(d, n^{1/d})$-LDSM and $(\bar{d}, n^{1/d})$-LDSM.
\end{remark}}

The relationship between LDSM with $L=n^{1/d}$, SM and DM is formulated in the following proposition.

\begin{proposition}\label{prop: LDSM DM SM connection}{\cite{vorobyev2021fast}}
\begin{enumerate}[label=\roman*)]
    \item For any natural $d$ satisfying $d \le n^{1/d}$ a $d$-DM is a $(\bar{d}, n^{1/d})$-LDSM. A $(\bar{d}, n^{1/d})$-LDSM is a $\bar{d}$-SM.
    \item For any natural $d$ satisfying $d \le n^{1/d}$ a $d$-DM is a $(d, n^{1/d})$-LDSM. A $(d, n^{1/d})$-LDSM is a $d$-SM.
\end{enumerate}
\end{proposition}

Denote the maximum number of columns of $d$-DM, $d$-SM, $\bar{d}$-SM, $d$-SSM, $(d, L)$-LDSM and $(\bar{d}, L)$-LDSM with $t$ rows as $N_{\text{DM}}(t, d)$, $N_{\text{SM}}(t, d)$, $N_{\text{SM}}(t, \bar{d})$, $N_{\text{SSM}}(t, d)$, $N_{\text{LDSM}}(t, d, L)$ and $N_{\text{LDSM}}(t, \bar{d}, L)$, respectively. 

For each family $\mathcal{F}\in\{\mathrm{DM},\mathrm{SM},\mathrm{SSM},\mathrm{LDSM}\}$ we define its \emph{largest rate} by
\begin{equation*}
    R_{\mathcal{F}}(\theta)
    = \limsup_{t\to\infty} \frac{1}{t}\log_2 N_{\mathcal{F}}(t,\theta),
\end{equation*}
where $\theta$ stands for $d$, $\bar d$, $(d,L)$, or $(\bar d,L)$ accordingly.

From Propositions \ref{prop: ssm sm dm connect} and \ref{prop: LDSM DM SM connection} it follows that
$$
\begin{aligned}
    R_{\text{DM}}(d) & \le R_{\text{SSM}}(d) \le R_{\text{SM}}(\bar{d}), \\
    R_{\text{DM}}(d) & \le R_{\text{LDSM}}(\bar{d}, n^{1/d}) \le R_{\text{SM}}(\bar{d}), \\
    R_{\text{DM}}(d) & \le R_{\text{LDSM}}(d, n^{1/d}) \le R_{\text{SM}}(d).
\end{aligned}
$$

\section{Results for \texorpdfstring{$d$}--SM and \texorpdfstring{$d, n^{1/d})$}--LDSM}\label{sec: SM LDSM lower bound}

For $d=2$ it is known~\cite{d1982bounds, dyachkov1989superimposed, coppersmith1998new, vorobyev2021fast, qian2022improved} {that}
$$
\begin{gathered}
0.1825 \le R_{\text{DM}}(2) \le 0.3219, \\
0.2237 \le R_{\text{SSM}}(2) \le 0.4998, \\
0.3135 \le R_{\text{SM}}(2) = R_{\text{SM}}(\bar{2}) \le 0.4998, \\
0.3017 \le R_{\text{LDSM}}(2, n^{1/2}) = R_{\text{LDSM}}(\bar{2}, n^{1/2}) \le 0.4998.
\end{gathered}
$$

One of the main results of our work is the following theorem, in which we present the lower bound on the largest rate of $d$-{SM} for arbitrary $d > 2$.

\begin{theorem}\label{th: d-SM lower bound}
{For arbitrary $d > 2$ the rate of $d$-SM can be lower bounded as
\begin{equation*}
R_{\text{SM}}(d) \ge \max_{p\in (0,1)} \min_{d_0, \alpha, \alpha_0, \alpha_1, \alpha_2} R^{(1)}, \nonumber
\end{equation*}
where the function $R^{(1)} = R^{(1)}(d, p, d_0, \alpha, \alpha_0, \alpha_1, \alpha_2)$ is defined as follows
\begin{equation*}
R^{(1)} = -\frac{1}{2d - d_0 - 1} \cdot \left(F_{bin, h} - 
\mathcal{A}^+_3\right),
\end{equation*}
$$
\begin{aligned}
F_{bin, h} 
& = h\left(\alpha\right) + \alpha\cdot h\left(\frac{\alpha_0}{\alpha}\right) + \alpha_0\cdot h\left(\frac{\alpha_1 + \alpha_0 - \alpha}{\alpha_0}\right) +  \\
& + \alpha_0\cdot h\left(\frac{\alpha_2 + \alpha_0 - \alpha}{\alpha_0}\right) - h\left(\alpha_0\right) - h\left(\alpha_1\right) - h\left(\alpha_2\right),
\end{aligned}
$$
$$
\begin{aligned}
\mathcal{A}^+_3 = A\left(d_0, p, \alpha_0\right) + A\left(d - d_0, p, \alpha_1\right) + A\left(d - d_0, p, \alpha_2\right),
\end{aligned}
$$
the function $A(s, Q, q)$ is defined in the parametric form
$$
\begin{aligned}
A(s, Q, q) 
& = (1-q) \cdot \log_2 (1-q) + q\cdot\log_2 \frac{Qy^s}{1-y} + \\
& + sQ \cdot \log_2 \frac{1-y}{y} + s\cdot h(Q),
\end{aligned}
$$
the parameter $y$, $0<y<1$, is defined as the unique root of the equation
\begin{equation*}
q = q(y) = Q\cdot\frac{1-y^s}{1-y},\; 0<y<1,
\end{equation*}
and the minimization is subject to the following constraints
\begin{equation*}
\alpha \in [p; \min(dp, 1)], \; \alpha_0 \in [\min(p, d_0p); \min(d_0p, \alpha)],
\end{equation*}
\begin{equation*}
\alpha_1, \alpha_2 \in [\max(\alpha-\alpha_0, p); \min((d-d_0)p, \alpha)], \; d_0 \in \{0,\dots, d-1\}.
\end{equation*}}
\end{theorem}

\textbf{Idea of the proof.} The main idea of the proof is that we build a $t\times n$ (with $n=\lfloor 2^{Rt} \rfloor$) matrix $C$, choosing each column of this matrix independently and uniformly from a fixed-weight column set with fixed weight $\lfloor pt \rfloor$, where $p \in (0, 1)$. {We show that with non-zero probability, this matrix is good enough and can be corrected to be a $d$-SM.} Then we find the optimal parameter $p$ to maximize the largest rate of this matrix.

This proof method differs from a typical random encoding method, in which each element of the matrix $C$ is selected from the set $\{0, 1\}$ with probability $p$. Usually, for such problems, the consideration of fixed-weight random matrices allows us to get better results than the random encoding method (see, for example,~\cite{dyachkov1989superimposed, johnson2018performance}). Indeed, the fixed weight column selection approach allows us to get an improved bound on the rate; however, it does come with a caveat of its own — the resulting expressions are much more challenging to manage. The detailed proof of Theorem \ref{th: d-SM lower bound} can be found in {Appendix \ref{sec: appendix a}}.

\begin{remark}
    {In the proof of Theorem \ref{th: d-SM lower bound}, we check that a random matrix satisfies the condition of Definition \ref{def: d-separable matrix} for the case of exactly $d$ defectives. In other words, we consider $2$ sets of columns $D_1$ and $D_2$ of size $d$ and estimate the probability of bad event $|\bigvee_{D_1} c_i| = |\bigvee_{D_2} c_i|$. The expression $R^{(1)} = R^{(1)}(p, d, d_0, \alpha, \alpha_0, \alpha_1, \alpha_2)$ is obtained by a combinatorial counting that is maximized over the following parameters: $|D_1 \cap D_2| = d_0$, $\frac{1}{t} \cdot |\bigvee_{D_1} c_i| = \frac{1}{t} \cdot |\bigvee_{D_2} c_i| = \alpha$, $\frac{1}{t} \cdot |\bigvee_{D_1\backslash D_2} c_i| = \alpha_1$, $\frac{1}{t} \cdot |\bigvee_{D_2\backslash D_1} c_i| = \alpha_2$, and $\frac{1}{t} \cdot |\bigvee_{D_1\cap D_2} c_i| = \alpha_0$.}
\end{remark}

{The lower bound presented in Theorem \ref{th: d-SM lower bound} coincides with lower bound presented in~\cite{d2014lectures} for small values of $d$. It is worth noting that in~\cite{d2014lectures} only numerical values are presented, and their proof is not published, while in this work we show an analytical formula. We present the obtained values of the lower bound on $d$-SM rate for small values of $d$ in Table \ref{tab: comparison}.}

{Remark that the main idea of the proof is that we build a random matrix $C$ and check whether this matrix is close to a $d$-SM. We can also check the second condition from the Definition \ref{def: d-LDSM} of the $(d, n^{1/d})$-LDSM and obtain the following result.}

\begin{theorem}\label{th: d-LDSM lower bound}
{In the notations of Theorem \ref{th: d-SM lower bound} the largest rate of $(d, n^{1/d})$-LDSM for $d>2$ can be lower bounded as
\begin{equation*}
R_{\text{LDSM}}(d, n^{1/d}) \ge \max_{p\in (0, 1)} \min \left(R^{(2)}, \min_{d_0, \alpha, \alpha_0, \alpha_1, \alpha_2} R^{(1)}\right), \nonumber
\end{equation*}
where the function $R^{(2)} = R^{(2)}(d, p)$ is defined as follows
\begin{equation*}
R^{(2)} = h(p) - dp\cdot h\left(\frac{1}{d} \right).
\end{equation*}}
\end{theorem}

{Note that expression $R^{(2)} = R^{(2)}(p, d)$ in the presented Theorem \ref{th: d-LDSM lower bound} appears due to additional condition on SM in Definition \ref{def: d-LDSM}. The detailed proof can be found in {Appendix \ref{sec: appendix a}}. We present the obtained numerical values of the lower bounds on $d$-SM and $(d, n^{1/d})$-LDSM largest rates for small values of $d$ in Table \ref{tab: comparison}. Note that for small values of $d > 2$, the resulting lower bounds coincide.}

\section{Results for \texorpdfstring{$\bar{d}$}--SM and \texorpdfstring{$(\bar{d}, n^{1/d})$}--LDSM}\label{sec: <= d SM LDSM}

{In this section, we discuss what results can be obtained for $\bar{d}$-SM and $(\bar{d}, n^{1/d})$-LDSM from the results for $d$-SM and $(d, n^{1/d})$-LDSM.} We formulate and prove two propositions. The first one is about the relationship between $\bar{d}$-SM, $d$-SM and $d$-DM. The second one is about the relationship between $(\bar{d}, n^{1/d})$-LDSM, $(d, n^{1/d})$-LDSM and $d$-DM. {The first of the presented propositions is not new in coding theory, but for the sake of completeness, we present its proof.} The detailed proofs of the following propositions are presented in {Appendix \ref{sec: appendix b}}.

\begin{proposition}\label{prop: <= and = SM}
A binary matrix $C$ is a $\bar{d}$-SM if and only if $C$ is a $d$-SM and a $(d-1)$-DM. 
\end{proposition}

\begin{proposition}\label{prop: <= and = LDSM}
A binary matrix $C$ is a $(\bar{d}, n^{1/d})$-LDSM if and only if $C$ is a $(d, n^{1/d})$-LDSM and a $(d-1)$-DM. 
\end{proposition}

\textbf{Idea of the results for $\bar{d}$-matrices.} {Recall that the main idea of the proof of Theorems \ref{th: d-SM lower bound}, \ref{th: d-LDSM lower bound} with lower bounds on the largest rates of $d$-SM and $(d, n^{1/d})$-LDSM is that we build a $t\times n$ (with $n=\lfloor 2^{Rt} \rfloor$) matrix $C$, choosing each column of this matrix independently and uniformly from a fixed-weight column set with weight $\lfloor pt \rfloor$, where $p \in (0, 1)$. Moreover, this method is used in the proof of the lower bound on the largest rate of $d$-DM~\cite{dyachkov1989superimposed}.} In both cases, the lower bounds have the form of a maximum concerning the parameter $p \in (0,1)$ of some function. For simplicity, we denote these lower bounds as follows
$$
\begin{aligned}
    R_{\text{DM}} (d) & \ge \max\limits_{p\in (0, 1)} \underline{R}_{\text{DM}} (d, p), \\
    R_{\text{SM}}(d) & \ge \max\limits_{p\in (0, 1)} \underline{R}_{\text{SM}} (d, p), \\
    R_{\text{LDSM}}(d, n^{1/d}) & \ge \max\limits_{p\in (0, 1)} \underline{R}_{\text{LDSM}} (d, n^{1/d}, p).
    \end{aligned}
$$

{Also denote optimal values of $p$ for $d$-DM, $d$-SM and $(d, n^{1/d})$-LDSM largest rate lower bounds as $p_{\text{DM}}(d)$, $p_{\text{SM}}(d)$ and $p_{\text{LDSM}}(d, n^{1/d})$, respectively.} Now we can formulate the main theorem of this section, the proof of which can be found in {Appendix \ref{sec: appendix b}}.

\begin{theorem}\label{th: SM LDSM lower bound <=d}
For arbitrary $d \ge 2$ the largest rates of $\bar{d}$-SM and $(\bar{d}, n^{1/d})$-LDSM can be lower bounded as
$$
\begin{aligned}
    R_{\text{SM}} & (\bar{d}) \ge \max\limits_{p \in (0, 1)} \left( \min \left(\underline{R}_{\text{DM}}(d-1, p), \underline{R}_{\text{SM}}(d, p)\right) \right), \\
    R_{\text{LDSM}}&(\bar{d}, n^{1/d}) \ge \\
    & \ge \max\limits_{p\in (0, 1)} \left( \min(\underline{R}_{\text{DM}}(d-1, p), \underline{R}_{\text{LDSM}}(d, n^{1/d}, p)) \right).
    \end{aligned}
$$
\end{theorem}

{To obtain the lower bounds on $\bar{d}$-SM and $(\bar{d}, n^{1/d})$-LDSM largest rates, presented in Table \ref{tab: comparison}, we can use the following inequalities
\begin{align*}
    & R_{\text{SM}}(\bar{d}) \ge \max\limits_{\begin{subarray}{l} p = p_{\text{DM}}(d-1),\\ p = p_{\text{SM}}(d) \end{subarray}} \left( \min(\underline{R}_{\text{DM}}(d-1, p), \underline{R}_{\text{SM}}(d, p)) \right), \\
    & R_{\text{LDSM}}(\bar{d}, n^{1/d}) \ge \\ 
    & \ge \max\limits_{\begin{subarray}{l} p = p_{\text{DM}}(d-1),\\ p = p_{\text{LDSM}}(d, n^{1/d}) \end{subarray}} \left( \min(\underline{R}_{\text{DM}}(d-1, p), \underline{R}_{\text{LDSM}}(d, n^{1/d}, p)) \right).
\end{align*}
}

We give the optimal values of the parameters $p_{\text{DM}}(d)$, $p_{\text{SM}}(d)$, $p_{\text{LDSM}}(d, n^{1/d})$, the optimal values of the lower bounds $\underline{R}_{\text{DM}} (d, p_{\text{DM}}(d))$, $\underline{R}_{\text{SM}} (d, p_{\text{SM}}(d))$, $\underline{R}_{\text{LDSM}} (d, n^{1/d}, p_{\text{LDSM}}(d, n^{1/d}))$, and the values of $\underline{R}_{\text{DM}} (d, p_{\text{SM}}(d + 1))$, $\underline{R}_{\text{DM}} (d, p_{\text{LDSM}}(d + 1, n^{1/d}))$, $\underline{R}_{\text{SM}} (d, p_{\text{DM}}(d - 1))$, $\underline{R}_{\text{LDSM}} (d, n^{1/d}, p_{\text{DM}}(d - 1))$ in Table \ref{tab: optimal params} in Appendix \ref{sec: appendix b} for small values of $d$.

\section{Conclusion} \label{sec: conclusion}

In this paper, we have discussed separable matrices (SM) and list-decoding separable matrices (LDSM) with list size $L=n^{1/d}$, which can identify a set of exactly $d$ or at most $d$ defectives with zero error probability among a set of $n$ elements using $t$ tests with decoding complexities $O(tn^d)$ for SM and $O(tn)$ for LDSM with $L=n^{1/d}$. We denote such matrices as $d$-SM, $\bar{d}$-SM, $(d, n^{1/d})$-LDSM and $(\bar{d}, n^{1/d})$-LDSM, respectively.

As one of the main contributions of the paper, we have presented and proved analytical lower bounds on the largest rates of $d$-SM for fixed $d \ge 3$. The presented lower bounds for $d$-SM coincide with numerical lower bounds previously published without proof in~\cite{d2014lectures}. Moreover, our results yield new lower bounds on the rate of $\bar{d}$-SM for $d \ge 3$.

This paper also presents new lower bounds on the rate of $(d, n^{1/d})$-LDSM and $(\bar{d}, n^{1/d})$-LDSM for fixed $d \ge 3$, as well as a comparison of these bounds with the best-known lower bounds on the rate of the $d$-SM, $\bar{d}$-SM, $d$-DM, and $d$-SSM. We have demonstrated that among matrices with efficient decoding complexity $O(tn)$, LDSM with $L=n^{1/d}$ have the best-known lower bounds on their largest rate. Moreover, for a small number of defectives $d \ge 3$, the presented lower bounds on the largest rates of SM and LDSM with $L=n^{1/d}$ coincide.

\newpage

\bibliographystyle{IEEEtran}
\bibliography{ref}

\newpage

\section{Appendix A}\label{sec: appendix a}
\subsection{Proof of Theorem \ref{th: d-SM lower bound}}
\begin{proof}
Consider a random binary matrix $C$ of size $t \times n$, where $n = \lfloor 2^{Rt} \rfloor$. Each column of this matrix is independently and uniformly chosen from a fixed-weight column set with length $t$ and weight $\lfloor pt \rfloor$, where $p \in (0, 1)$. In the following discussion, we will omit the $\lfloor. \rfloor$ notation for simplicity, as it does not affect the asymptotic rate. Consider two sets of indices $D_1$ and $D_2$, where $D_1$ and $D_2$ are subsets of $[n]$, and $|D_1| = |D_2| = d$. We call a pair of sets $D_1$ and $D_2$ \textit{bad}, if
\begin{equation}
\bigvee_{i\in D_1} c_i=\bigvee_{i\in D_2} c_i.
\end{equation}

\begin{remark}
{For simplicity of further presentation, we provide a proof plan.
\begin{enumerate}[label=\roman*)]
    \item Our first goal is to estimate the expected value $E$ of the number of bad pairs $D_1$ and $D_2$ for any matrix $C$.
    \item Then we estimate the maximal value $R = R_{max}$ such that $E < n/4$. For such a value of $R_{max}$, by Markov’s inequality, there is at most a $1/2$ probability that the number of bad pairs is $n/2$ or greater.
    \item We conclude that for big enough $t$, there exists a $t \times n$ matrix $C$ with a rate $R_{max}$ and the property that the expected number of bad pairs is at most $n/2$. By removing one column from each bad pair, we obtain a $d$-SM matrix with $t$ rows, at least $n/2$ columns and largest rate $R_{max}$. Therefore, $R_{SM}(d) \ge \limsup\limits_{t\to\infty} R_{max}$.
\end{enumerate}
}
\end{remark}

\textbf{Estimation of the expected number of bad pairs.} Let us estimate the expected value $E$ of the number of bad pairs of subsets. To do this, we introduce some notations. We denote $|\bigvee_{D_1} c_i|$ and $|\bigvee_{D_2} c_i|$ as $k$, which is the weight of a possible outcome vector $\mathbf{r}$. Similarly, we denote $|\bigvee_{D_1\backslash D_2} c_i| = k_1$, $|\bigvee_{D_2\backslash D_1} c_i| = k_2$, and $|\bigvee_{D_1\cap D_2} c_i| = k_0$.

The probability $P_{d_0}$ of a bad pair for two subsets $|D_1|$ and $|D_2|$ such that $|D_1 \cap D_2| = d_0$ can be written as
\begin{equation}\label{eq: p_d_0}
    P_{d_0} = \sum\limits_{k} \sum\limits_{k_0} \sum\limits_{k_1} \sum\limits_{k_2} F(p, d, d_0, k, k_0, k_1, k_2),
\end{equation}
where $F = F(p, d, d_0, k, k_0, k_1, k_2)$ is defined as
\begin{equation}
     F \overset{\Delta}{=} \frac{\binom{t}{k} \cdot \left( \binom{k}{k_0} \cdot \frac{P^{(t)}(d_0, p, k_0)}{\binom{t}{k_0}} \cdot \binom{t}{pt}^{d_0} \right) \cdot F_1(k_1) \cdot F_1(k_2)}{\binom{t}{pt}^{2d-d_0}}
\end{equation}
the expression $P^{(t)}(d, p, k_0)$ is the probability that the Boolean sum of $d$ columns of weight $pt$ is a column of weight $k_0$, and the expression $F_1 (k_i)$ is defined as follows
\begin{equation*}
    F_1 (k_i) \overset{\Delta}{=} \binom{k_0}{k_i-(k-k_0)} \cdot \frac{P^{(t)}(d - d_0, p, k_i)}{\binom{t}{k_i}} \cdot \binom{t}{pt}^{d-d_0}.
\end{equation*}
The summation takes place according to the following parameters
\begin{equation*}
    k\in \{pt, \dots, \min(dpt, t)\},
\end{equation*}
\begin{equation*}
    k_0\in\{\min(pt, d_0pt), \dots, \min(d_0pt, k)\},
\end{equation*}
\begin{equation*}
    k_1, k_2 \in \{\max(k-k_0, pt), \dots, \min((d-d_0)pt, k))\}.
\end{equation*}

The main instruments for further transformations are Stirling's formula for binomial coefficients and asymptotics of $P^{(t)}(d,p,k)$, presented in~\cite{d2015almost}. Our next goal is to estimate the mathematical expectation $E_{d_0}$ of the number of bad pairs with $d_0$ intersection cardinality. The expression $E_{d_0}$, its logarithm and logarithm of $P_{d_0}$ can be upper bounded as follows
\begin{equation*}
E_{d_0} \le n^{2d-d_0} \cdot P_{d_0},
\end{equation*}
\begin{equation}\label{eq: log E_d0}
\log_2 E_{d_0} \le\left(2 d-d_0\right) \log_2 n+\log_2 P_{d_0},
\end{equation}
\begin{equation}\label{eq: log P_d0}
\log_2 P_{d_0} \le \log_2 \left(t^4 \max_{k, k_0, k_1, k_2} F\right) = o\left(\log_2 n\right) + \max_{k, k_0, k_1, k_2}\left(\log_2 F\right).
\end{equation}

{Then we can rewrite $F$ as a product of two expressions $F = F_{bin} \cdot F_{pr}$, where
\begin{equation}\label{eq: F product}
    \begin{aligned}
    F_{bin} & \overset{\Delta}{=} \frac{\binom{t}{k}\cdot \binom{k}{k_0}}{\binom{t}{k_0}} \cdot \frac{\binom{k_0}{k_1-k+k_0}}{\binom{t}{k_1}} \cdot \frac{\binom{k_0}{k_2-k+k_0}}{\binom{t}{k_2}} \\
    F_{pr} & \overset{\Delta}{=} P^{(t)}(d_0, p, k_0) \cdot P^{(t)}(d-d_0, p, k_1) \cdot P^{(t)}(d-d_0, p, k_2).
    \end{aligned}
\end{equation}}

Using the {Stirling's approximation} $\binom{n}{k} = 2^{n\cdot\left(h\left(\frac{k}{n}\right) + o(1)\right)}$ and replacing $k, k_0, k_1, k_2$ with $\alpha, \alpha_0, \alpha_1, \alpha_2$ by dividing {these parameters} by $t$, we can rewrite $\log_2 F_{bin}$ as $\log_2 F_{bin} = t\cdot F_{bin,h}$
where 
\begin{align}\label{eq: F_bin}
    F_{bin,h} 
    & \overset{\Delta}{=} h\left(\alpha\right) + \alpha\cdot h\left(\frac{\alpha_0}{\alpha}\right) - h(\alpha_0) + \alpha_0\cdot h\left(\frac{\alpha_1 + \alpha_0 - \alpha}{\alpha_0}\right) - \nonumber\\
    & - h\left(\alpha_1\right) + \alpha_0\cdot h\left(\frac{\alpha_2 + \alpha_0 - \alpha}{\alpha_0}\right) - h\left(\alpha_2\right) + o(1).
\end{align}

Now we need to estimate the expressions $F_{pr}$ and $\log_2 F_{pr}$. By~\cite{d2015almost} it is known that
\begin{equation*}
P^{(N)}(s, Q, k) \overset{\Delta}{=} Pr(|\vee_s c_i| = k), \; QN\le k \le min(N, sQN),
\end{equation*}
\begin{equation*}
A(s, Q, q) \overset{\Delta}{=} \lim\limits_{N\to\infty} \frac{-\log_2 P^{(N)}(s, Q, qN)}{N},
\end{equation*}
$$
    \begin{aligned}
    A(s, Q, q) 
    & = (1-q) \cdot \log_2 (1-q) + q\cdot\log_2 \frac{Qy^s}{1-y} + \\ 
    & + sQ \cdot \log_2 \frac{1-y}{y} + s\cdot h(Q),
    \end{aligned}
$$
where the parameter $y$ is defined as a unique solution of the following equation
\begin{equation*}
q = q(y) = Q\cdot\frac{1-y^s}{1-y},\; 0<y<1.
\end{equation*}

Then $\log_2 F_{pr}$ can be written as follows
\begin{align}\label{eq: F_pr}
& \log_2 F_{pr} = \\ 
& = t\cdot (-A(d_0, p, \alpha_0) - A(d - d_0, p, \alpha_1) - A(d - d_0, p, \alpha_2) + o(1)). \nonumber
\end{align}

Return to the estimation of mathematical expectation $E_{d_0}$ and substitute \eqref{eq: log P_d0}, \eqref{eq: F product}, \eqref{eq: F_bin}, \eqref{eq: F_pr} into \eqref{eq: log E_d0}:
$$
\begin{aligned}
\log_2 E_{d_0} 
& \le  \left(2 d-d_0\right) \log_2 n+\log_2 P_{d_0} \le \\
& \le \left(2 d-d_0\right) \cdot \log_2 n+ o\left(\log_2 n\right) + \max_{k, k_0, k_1, k_2}\left(\log_2 F\right) = \\
& = \left(2d - d_0 + o(1)\right) \cdot \log_2 n + t\cdot \max_{\alpha, \alpha_0, \alpha_1, \alpha_2} H_{d_0},
\end{aligned}
$$
where 
\begin{equation}
H_{d_0} \overset{\Delta}{=} F_{bin,h} - A(d_0, p, \alpha_0) - A(d - d_0, p, \alpha_1) - A(d - d_0, p, \alpha_2).
\end{equation}

\textbf{Finding the maximum appropriate value of $R$.} {Define $R_{max}$ as the maximum value $R$ such that $E_{d_0} < n/4d$ for every $d_0$. For such a value of $R = R_{max}$, the expected value of the number of bad pairs is less than $n/4$. Hence, by Markov’s inequality, there is at most a $1/2$ probability that the number of bad pairs is $n/2$ or greater. Let us estimate this parameter $R_{max}$. We demand that for every $d_0$, the following inequality holds}
\begin{equation*}
\left(2 d-d_0 + o(1)\right) \log_2 n + t \max_{\alpha, \alpha_0, \alpha_1, \alpha_2} H_{d_0} \le \log_2 n + o (\log n),
\end{equation*}
\begin{equation}
R = \frac{\log_2 n}{t} \le \frac{1}{2d - d_0 - 1} \cdot \min_{\alpha, \alpha_0, \alpha_1, \alpha_2} (-H_{d_0}) + o(1).
\end{equation}

To find the optimal value of $R$, we can maximize the right part of the resulting upper bound by $p$ and select the largest $R$ satisfying the resulting inequality. Therefore, 
\begin{equation}
\begin{aligned}
    R_{max} 
    & \ge \max_{p} \min_{d_0, \alpha, \alpha_0, \alpha_1, \alpha_2} \left(-\frac{H_{d_0}}{2d - d_0 - 1}\right) + o(1) = \\
    & = \max_{p} \min_{d_0, \alpha, \alpha_0, \alpha_1, \alpha_2} \left(R^{(1)}\right) + o(1).
\end{aligned}
\end{equation}

\textbf{Correction of a random matrix.} From the presented arguments it follows that for big enough $t$, there exists a $t \times n$ matrix $C$ with a rate $R = \max\limits_{p} \min\limits_{d_0, \alpha, \alpha_0, \alpha_1, \alpha_2} R^{(1)} + o(1)$ with the following property: the expected number of bad pairs is at most $n/2$. By removing one column from each bad pair, we obtain a $d$-SM matrix with $t$ rows, at least $n/2$ columns and largest rate $\max\limits_{p} \min\limits_{d_0, \alpha, \alpha_0, \alpha_1, \alpha_2} R^{(1)} + o(1)$. Therefore, the largest rate of $d$-SM satisfies the following inequality 
\begin{equation}
    R_{SM}(d) \ge \max\limits_{p} \min\limits_{d_0, \alpha, \alpha_0, \alpha_1, \alpha_2} R^{(1)}.    
\end{equation}
\end{proof}

\subsection{Proof of Theorem \ref{th: d-LDSM lower bound}}
\begin{proof}
    {The proof of this theorem is a continuation of the proof of Theorem \ref{th: d-SM lower bound}. Again, we consider a random binary matrix $C$ of size $t \times n$, where $n = \lfloor 2^{Rt} \rfloor$. Each column of this matrix is independently and uniformly chosen from a fixed-weight column set with length $t$ and weight $\lfloor pt \rfloor$, where $p \in (0, 1)$. In Theorem \ref{th: d-SM lower bound}, we have checked that for such a matrix and quite big $t$, there is at most $1/2$ probability that the number of bad pairs of $ d$-column subsets is at least $n/2$.}

    Our next goal is to {estimate} the maximal value $R = \hat{R}_{max}$ for which the second condition from Definition \ref{def: d-LDSM} holds with $L=n^{1/d}$. We bound the probability $P$ that one fixed vector of weight $\le dpt$ covers more than $\sqrt[d]{\frac{n}{2}}$ columns{, and show that $2^t \cdot P \to 0$ with $n \to \infty$. In other words, we show that the probability that at least one vector of weight $\le dpt$ covers more than $\sqrt[d]{\frac{n}{2}}$ columns, tends to $0$.} The number of covered columns is stochastically dominated by a binomial random variable $\xi \sim Bin(n, q)$, where $q$ is the probability that one fixed vector of weight $dpt$ covers one column of weight $pt$. This probability can be written as follows
    \begin{equation}\label{eq: q}
    \begin{aligned}
        q 
        & = \frac{\binom{dpt}{pt}}{\binom{t}{pt}} = \frac{2^{dpt\left(h\left(\frac{1}{d}\right) + o(1)\right)}}{2^{t(h(p) + o(1))}} = 2^{t\cdot\left(dp\cdot h\left(\frac{1}{d}\right) + o(1) - h(p)\right)} = \\
        & = n^{\left(dp\cdot h\left(\frac{1}{d}\right) + o(1) - h(p)\right)/R} = n^{\beta(p, d) + o(1)},
    \end{aligned}
    \end{equation}
    {where $\beta(p, d) = \frac{dp\cdot h\left(\frac{1}{d}\right) - h(p)}{R}$. We require that for some small $\epsilon > 0$ the following inequality holds $\beta(p, d) < -1 - \epsilon$. We can rewrite the last inequality in the following form $R < \frac{h(p) - dp\cdot h\left(\frac{1}{d}\right)}{1 + \epsilon}$.}
    
    The probability $P$ can be upper bounded as follows
    \begin{equation}\label{eq: P}
    P \le \sum\limits^{n}_{k=\sqrt[d]{\frac{n}{2}}} \binom{n}{k}\cdot q^k\cdot (1-q)^{n-k}.
    \end{equation}
    For the binomial distribution, $\lfloor (n+1)\cdot q \rfloor$ is the mode, that is, the value with the highest probability in $Bin(n,q)$. At the same time, $\lfloor (n+1)\cdot q \rfloor \le (n + 1) \cdot q = n^{\beta(p, d) + o(1)} + n^{1 + \beta(p, d) + o(1)} < 2\cdot n^{1 + \beta(p, d) + o(1)} < 2 < (n/2)^{\frac{1}{d}}$ for big enough $n$. Then the largest term of the sum of the upper bound on $P$ is the first one. Using the equalities \eqref{eq: q} and \eqref{eq: P} we obtain that
    
    \begin{align}\label{eq: P next}
    P 
    & \le n\cdot \binom{n}{\sqrt[d]{\frac{n}{2}}} \cdot q^{\sqrt[d]{\frac{n}{2}}} = \frac{n\cdot n! \cdot n^{(\beta(p, d) + o(1)) \cdot \sqrt[d]{\frac{n}{2}}}}{\left(\sqrt[d]{\frac{n}{2}}\right)! \cdot \left(n - \sqrt[d]{\frac{n}{2}}\right)!} \le \nonumber \\ 
    & \le n\cdot n^{\sqrt[d]{\frac{n}{2}}} \cdot n^{(\beta(p, d) + o(1)) \cdot \sqrt[d]{\frac{n}{2}}} \le n^{(1 + \beta(p, d) + o(1)) \cdot \sqrt[d]{\frac{n}{2}}}.
    \end{align}
    
    {The probability that at least one vector of weight not greater than $dpt$ covers more than $\sqrt[d]{\frac{n}{2}}$ columns can be upper bounded as $2^t\cdot P \le n^{\frac{1}{R} + (1 + \beta(p, d) + o(1))\cdot \sqrt[d]{\frac{n}{2}}}.$ Since $1 + \beta(p, d) + o(1) < -\epsilon + o(1) < -\epsilon/2$ for big enough $t$, then $2^t\cdot P \rightarrow_n 0$.} {Tending $\epsilon \to 0$ we obtain that $R \le R^{(2)} + o(1)$, where $R^{(2)} \overset{\Delta}{=} h(p) - dp \cdot h \left(\frac{1}{d}\right)$.}
    
    {We have estimated the probabilities that random matrix $C$ has more than $n/2$ bad pairs of $d$ column subsets and that one fixed vector of weight $\le dpt$ covers more than $\sqrt[d]{\frac{n}{2}}$ columns of $C$. Then we can estimate the probability of the union of these bad events as $1/2 + o(1) < 1$ for big enough $t$. We also find bounds for $R$ at which probability estimates hold. Then both probability estimates are satisfied for $R \le \max\limits_{p\in (0, 1)} \min \left(R^{(2)}, \min\limits_{d_0, \alpha, \alpha_0, \alpha_1, \alpha_2} R^{(1)}\right) + o(1)$.}
    
    {We can conclude that for big enough $t$, there exists a $t \times n$ matrix $C$ with the following properties: the expected number of bad pairs is at most $n/2$, and the number of columns that can be covered by a vector of weight $\le dpt$ is at most $\sqrt[d]{\frac{n}{2}}$. By removing one column from each bad pair, we obtain a $(d, n^{1/d})$-LDSM with $t$ rows, not less than $n/2$ columns and largest rate $\max\limits_{p\in (0, 1)} \min \left(R^{(2)}, \min\limits_{d_0, \alpha, \alpha_0, \alpha_1, \alpha_2} R^{(1)}\right)$. Therefore, 
    \begin{equation}
        R_{LDSM}(d, n^{1/d}) \ge \max_{p\in (0, 1)} \min \left(R^{(2)}, \min_{d_0, \alpha, \alpha_0, \alpha_1, \alpha_2} R^{(1)}\right).
    \end{equation}}
\end{proof}

Using numerical calculations, we obtain the results, shown in Table \ref{tab: results2}.

\begin{table}[htbp]
  \centering
    \caption{Lower bound on the largest rate of $(d, n^{1/d})$-LDSM and the corresponding parameters for small $d$.}
    \begin{tabular}{| c | c | c | c | c | c |}
        \hline
        $d$ & 2 & 3 & 4 & 5 & 6 \\ 
        \hline
        $p$ & 0.31 & 0.22 & 0.17 & 0.14 & 0.11 \\
        \hline
        $\beta(p, d)$ & -0.904 & -1.085 & -1.293 & -1.488 & -1.916 \\
        \hline
        $\underline{R}_{\text{LDSM}}(d, n^{1/d})$ & 0.302 & 0.142 & 0.082 & 0.053 & 0.037\\
        \hline
    \end{tabular}
    \label{tab: results2}
\end{table}

\section{Appendix B}\label{sec: appendix b}
\subsection{Proof of Proposition \ref{prop: <= and = SM}}
\begin{proof}
"$\Rightarrow$" Let $C$ be a $\bar{d}$-SM. Then by Definition \ref{def: d-LDSM} for any two different sets $D_1, D_2 \subset [n]$, $|D_1| \le d$, $|D_2| \le d$, the Boolean sums $\bigvee _{i\in D_1} c_i$ and $\bigvee_{i\in D_2} c_i$ are distinct. But then the condition $\bigvee _{i\in D_1} c_i \neq \bigvee_{i\in D_2} c_i$ is also true with $|D_1| = |D_2| = d$. Therefore, $C$ is a $d$-SM.

From Proposition \ref{prop: SM DM connectness} it follows that a $\bar{d}$-SM is a $(d-1)$-DM. So $C$ is a $(d-1)$-DM, too.

"$\Leftarrow$" Let $C$ be a $d$-SM and a $(d-1)$-DM. Note that from Proposition \ref{prop: SM DM connectness} it follows that $(d-1)$-DM is a $(\overline{d-1})$-SM, so $C$ is a $(\overline{d-1})$-SM. We need to check the condition from Definition \ref{def: d-separable matrix} for $C$ to be a $\bar{d}$-SM. 

Consider two different sets $D_1, D_2 \subset [n]$, $|D_1| \le d$, $|D_2| \le d$. The Boolean sums $\bigvee _{i\in D_1} c_i$ and $\bigvee_{i\in D_2} c_i$ are distinct when $|D_1| = |D_2| = d$, since $C$ is a $d$-SM. Such Boolean sums are also distinct when $|D_1| < d$ and $|D_2| < d$, since $C$ is a $(\overline{d-1})$-SM. The last case is when $|D_1| = d$ and $|D_2| < d$. Then there exist a vector $\hat{c} \in \{c_i, i\in D_1\}$ and $\hat{c} \notin \{c_i, i\in D_2\}$. If $\bigvee _{i\in D_1} c_i = \bigvee_{i\in D_2} c_i$, then the Boolean sum $\bigvee_{i\in D_2} c_i$ covers $\hat{c}$. This is a contradiction with Definition \ref{def: d-disjunctive matrix} of a $(d-1)$-DM. Therefore, $C$ is a $\bar{d}$-SM.
\end{proof}

\subsection{Proof of Proposition \ref{prop: <= and = LDSM}}
\begin{proof}
"$\Rightarrow$" Let $C$ be a $(\bar{d}, n^{1/d})$-LDSM. Then by Definition \ref{def: d-LDSM} matrix $C$ is a $\bar{d}$-SM and by Proposition \ref{prop: <= and = SM} matrix $C$ is a $d$-SM and a $(d-1)$-DM. By Definition \ref{def: d-LDSM}, each vector, which can be obtained as a Boolean sum of \textit{at most} $d$ columns of $C$, covers at most $n^{1 / d}$ columns from $C$. Therefore, each vector, which can be obtained as a Boolean sum of \textit{exactly} $d$ columns of $C$, covers at most $n^{1 / d}$ columns from $C$. Therefore, $C$ is a $(d, n^{1/d})$-LDSM and a $(d-1)$-DM.

"$\Leftarrow$" Let $C$ be a $(d, n^{1/d})$-LDSM and a $(d-1)$-DM. Then by Definition \ref{def: d-LDSM} matrix $C$ is a $d$-SM, and by Proposition \ref{prop: <= and = SM} matrix $C$ is a $\bar{d}$-SM. By Definition \ref{def: d-LDSM}, each vector, which can be obtained as a Boolean sum of \textit{exactly} $d$ columns of $C$, covers at most $n^{1 / d}$ columns from $C$. But this means that a smaller number of columns (less than $d$) also can not cover more than $n^{1 / d}$ columns from $C$. Therefore, $C$ is a $(\bar{d}, n^{1/d})$-LDSM.
\end{proof}

\subsection{Proof of Theorem \ref{th: SM LDSM lower bound <=d}}
\begin{proof}
\textbf{Separable matrices.} By Proposition \ref{prop: <= and = SM} we can build a $\bar{d}$-SM as a matrix $C$ satisfying two conditions: 
\begin{itemize}
    \item Matrix $C$ is a $d$-SM.
    \item Matrix $C$ is a $(d-1)$-DM.
\end{itemize}

As discussed earlier, proofs of the lower bounds on the $d$-DM largest rate~\cite{dyachkov1989superimposed} and $d$-SM largest rate [this work] use the same method of the proof. We build a $t \times n$ matrix $C$ and choose each column of this matrix independently and uniformly from a fixed-weight column set with weight $\lfloor pt \rfloor$, where $p \in (0, 1)$.

Therefore, we can build a matrix $C$ from $\bar{d}$-SM family using the same method and, when building, check for which $R$ both described conditions are satisfied. Then for any fixed $p \in (0, 1)$ the rate of the built matrix $R_{\text{SM}} (\bar{d}, p)$ can be lower bounded as follows
\begin{equation}
    R_{\text{SM}} (\bar{d}, p) \ge \min \left(\underline{R}_{\text{D}}(d-1,p), \underline{R}_{\text{SM}}(d,p)\right).
\end{equation}

Therefore, $R_{\text{SM}}(\bar{d})$ can be lower bounded as
\begin{equation}
    R_{\text{SM}}(\bar{d}) \ge \max\limits_{p \in (0, 1)} \left( \min \left(\underline{R}_{\text{D}}(d-1,p), \underline{R}_{\text{SM}}(d,p)\right) \right).
\end{equation}

\textbf{List-decoding separable matrices with $L=n^{1/d}$.}
{We can use the same idea as for the SM family. By Proposition \ref{prop: <= and = LDSM} we can build a $(\bar{d}, n^{1/d})$-LDSM as a matrix $C$ satisfying two conditions: 
\begin{itemize}
    \item Matrix $C$ is a $(d, n^{1/d})$-LDSM.
    \item Matrix $C$ is a $(d-1)$-DM.
\end{itemize}}

Then we build a $t \times n$ matrix $C$ and choose each column of this matrix independently and uniformly from a fixed-weight column set with weight $\lfloor pt \rfloor$, where $p \in (0, 1)$. When building, we check for which $R$ both conditions are satisfied. Then for any fixed $p \in (0, 1)$ the rate of the built matrix is $R_{\text{LDSM}} (\bar{d}, n^{1/d}, p)$ can be lower bounded as follows
\begin{equation}
    R_{\text{LDSM}} (\bar{d}, n^{1/d}, p) \ge \min \left(\underline{R}_{\text{D}}(d-1,p), \underline{R}_{\text{LDSM}}(d, n^{1/d},p)\right).
\end{equation}

Therefore, $R_{\text{LDSM}}(\bar{d}, n^{1/d})$ can be lower bounded as
\begin{equation}
    R_{\text{LDSM}}(\bar{d}, n^{1/d}) \ge \max\limits_{p \in (0, 1)} \left( \min \left(\underline{R}_{\text{D}}(d-1,p), \underline{R}_{\text{LDSM}}(d, n^{1/d},p)\right) \right).
\end{equation}
\end{proof}

\begin{table}[htbp]
  \centering
    \caption{Optimal parameters for $d$-SM, $(d, n^{1/d})$-LDSM and $d$-DM.}
    \begin{tabular}{| c | c | c | c | c | c |}
        \hline
        $d$ & 2 & 3 & 4 & 5 & 6 \\ 
        \hline \hline
        $p_{\text{DM}}(d)$~\cite{dyachkov1989superimposed} & 0.26 & 0.19 & 0.15 & 0.12 & 0.10 \\
        \hline
        $p_{\text{SM}}(d)$ [this work] & 0.31 & 0.22 & 0.17 & 0.14 & 0.11 \\ 
        \hline
        $p_{\text{LDSM}}(d, n^{1/d})$ [this work] & 0.31 & 0.22 & 0.17 & 0.14 & 0.11 \\
        \hline \hline
        $\underline{R}_{\text{DM}} (d, p_{\text{DM}}(d))$ & 0.183 & 0.079 & 0.044 & 0.028 & 0.019 \\
        \hline 
        $\underline{R}_{\text{SM}} (d, p_{\text{SM}} (d))$ & 0.302 & 0.142 & 0.082 & 0.053 & 0.037 \\
        \hline 
        $\underline{R}_{\text{LDSM}} (d, n^{1/d}, p_{\text{LDSM}} (d, n^{1/d}))$ & 0.302 & 0.142 & 0.082 & 0.053 & 0.037 \\
        \hline \hline
        $\underline{R}_{\text{DM}} (d, p_{\text{SM}}(d+1))$ & 0.180 & 0.078 & 0.044 & 0.028 & 0.019 \\
        \hline
        $\underline{R}_{\text{DM}} (d, p_{\text{LDSM}}(d+1, n^{1/(d+1)}))$ & 0.180 & 0.078 & 0.044 & 0.028 & 0.019 \\
        \hline
        $\underline{R}_{\text{SM}} (d, p_{\text{DM}}(d-1))$ & 1.000 & 0.140 & 0.081 & 0.053 & 0.037 \\
        \hline
        $\underline{R}_{\text{LDSM}} (d, n^{1/d}, p_{\text{DM}}(d-1))$ & 1.000 & 0.140 & 0.081 & 0.053 & 0.037 \\
        \hline
    \end{tabular}
    \label{tab: optimal params}
\end{table}

\end{document}